\documentstyle[twoside,epsfig]{article}                                                                    
\def\3{\ss}                                                                                        

\newcommand{\qq}         {\mbox{$Q^2$}}
\newcommand{\xjet}       {\mbox{$x_{Jet}$}}
\newcommand{\etajet}     {\mbox{$\eta_{Jet}$}}
\newcommand{\ETQQ}       {\mbox{$E^2_{T,jet}/Q^2$}}

\begin{document}

\title {
        \begin{flushright}{\normalsize DESY--99--162}\end{flushright}
        \vspace{2cm}
        \bf\LARGE  Measurement of the $E_{T,jet}^2/Q^2$ dependence \\
                   of forward--jet production at HERA}
        \vspace{1cm}
        \author{ZEUS Collaboration}  
        \date{}

\maketitle

\vspace*{1cm}

\begin{abstract}
\noindent
The forward--jet cross section in deep inelastic $e^{+}p$ scattering has
been measured using the
ZEUS detector at HERA with an integrated luminosity of 6.36~pb$^{-1}$.
The jet cross section is presented as a function of 
jet transverse energy squared, $E^2_{T,jet}$, and $Q^2$
in the kinematic ranges $10^{-2}<E^2_{T,jet}/Q^2<10^2$
and $2.5\cdot10^{-4}<x<8.0\cdot 10^{-2}$.
Since the perturbative QCD predictions for this cross section
are sensitive to the treatment of the $\log(\ETQQ)$ terms, this
measurement provides an important test.
The measured cross section is compared to the predictions
of a next--to--leading order pQCD calculation as well as to 
various leading--order Monte Carlo models.
Whereas the predictions of all models agree with the measured cross section in
the region of small 
$E^2_{T,Jet}/Q^2$, only one model, which
includes a resolved photon component, describes the data
over the whole kinematic range.  
\end{abstract}
\pagestyle{plain}                   
\thispagestyle{empty}
\newpage                                                     
%
\pagenumbering{Roman}                                                                              

\begin{center}                                                                                     
{                      \Large  The ZEUS Collaboration              }                               
\end{center}                                                                                       
  J.~Breitweg,                                                                                     
  S.~Chekanov,                                                                                     
  M.~Derrick,                                                                                      
  D.~Krakauer,                                                                                     
  S.~Magill,                                                                                       
  B.~Musgrave,                                                                                     
  A.~Pellegrino,                                                                                   
  J.~Repond,                                                                                       
  R.~Stanek,                                                                                       
  R.~Yoshida\\                                                                                     
 {\it Argonne National Laboratory, Argonne, IL, USA}~$^{p}$
% ...
\newpage                                                                                           
$^{\    1}$ now visiting scientist at DESY \\                                                      
% ...
                                                           %
                                                           %
\newpage   
                                                           %
                                                           %
\begin{tabular}[h]{rp{14cm}}                                                                       
$^{a}$ &  supported by the Natural Sciences and Engineering Research                               
          Council of Canada (NSERC)  \\                                                            
$^{b}$ &  supported by the FCAR of Qu\'ebec, Canada  \\                                            
$^{c}$ &  supported by the German Federal Ministry for Education and                               
          Science, Research and Technology (BMBF), under contract                                  
          numbers 057BN19P, 057FR19P, 057HH19P, 057HH29P, 057SI75I \\                              
$^{d}$ &  supported by the MINERVA Gesellschaft f\"ur Forschung GmbH, the                          
German Israeli Foundation, and by the Israel Ministry of Science \\                                
$^{e}$ &  supported by the German-Israeli Foundation, the Israel Science                           
          Foundation, the U.S.-Israel Binational Science Foundation, and by                        
          the Israel Ministry of Science \\                                                        
$^{f}$ &  supported by the Italian National Institute for Nuclear Physics                          
          (INFN) \\                                                                                
$^{g}$ &  supported by the Japanese Ministry of Education, Science and                             
          Culture (the Monbusho) and its grants for Scientific Research \\                         
$^{h}$ &  supported by the Korean Ministry of Education and Korea Science                          
          and Engineering Foundation  \\                                                           
$^{i}$ &  supported by the Netherlands Foundation for Research on                                  
          Matter (FOM) \\                                                                          
$^{j}$ &  supported by the Polish State Committee for Scientific Research,                         
          grant No. 115/E-343/SPUB/P03/154/98, 2P03B03216, 2P03B04616,                             
          2P03B10412, 2P03B03517, and by the German Federal                                        
          Ministry of Education and Science, Research and Technology (BMBF) \\                     
$^{k}$ &  supported by the Polish State Committee for Scientific                                   
          Research (grant No. 2P03B08614 and 2P03B06116) \\                                        
$^{l}$ &  partially supported by the German Federal Ministry for                                   
          Education and Science, Research and Technology (BMBF)  \\                                
$^{m}$ &  supported by the Fund for Fundamental Research of Russian Ministry                       
          for Science and Edu\-cation and by the German Federal Ministry for                       
          Education and Science, Research and Technology (BMBF) \\                                 
$^{n}$ &  supported by the Spanish Ministry of Education                                           
          and Science through funds provided by CICYT \\                                           
$^{o}$ &  supported by the Particle Physics and                                                    
          Astronomy Research Council \\                                                            
$^{p}$ &  supported by the US Department of Energy \\                                              
$^{q}$ &  supported by the US National Science Foundation                                          
\end{tabular}                                                                                      

\newpage
\pagenumbering{arabic}
\section{Introduction}
The wide kinematic range available at the HERA $ep$ collider
at DESY has allowed QCD to be tested in regions
of phase space not available to previous experiments.
Both the H1 and ZEUS collaborations have studied the forward--jet 
cross sections~\cite{H1_forward,ZEUS_FJets} in order to  search for BFKL
effects~\cite{BFKL,MUELLER91}.
For these analyses, the two hard scales involved in jet production in 
deep inelastic scattering (DIS), 
the negative square of the four--momentum transfer at the lepton
vertex, $Q^2$, and the squared transverse jet energy, $E_{T,jet}^2$, 
were chosen to be of the same order of magnitude.
This paper extends our previous study~\cite{ZEUS_FJets}  
by investigating the forward--jet cross section as a function of the ratio 
of these two scales, $E^2_{T,jet}/Q^2$, for the entire available range.\\

%
Three different kinematic regions can be distinguished,
depending on the dominant scale. 
In the first region, $Q^2 \gg E^2_{T,jet}$, $Q^{2}$ is the standard 
deep inelastic process hard scale. Typically, leading--order (LO) 
Monte Carlo models 
approximate pQCD contributions in this regime by parton showers.
In the second region, where $E^2_{T,jet} \approx Q^2$, all terms with
$\log(Q^2/E^2_{T,jet})$ become small and the effects of 
DGLAP evolution~\cite{DGLAP}
are suppressed. Therefore BFKL effects are expected to 
be observable in this region, which was selected for the analysis
of forward--jet production~\cite{ZEUS_FJets}, where it was
discussed in detail.
In the third region, where $Q^2 \ll E^2_{T,jet}$, 
the NLO pQCD prediction is sensitive to the treatment of terms proportional 
to $\log(E^2_{T,jet}/Q^2)$, which ought to be resummed.
Conventional Monte Carlo models do not include these terms.\\

In this letter,
measurements of the forward--jet cross sections covering all three regions 
are presented and compared to the predictions of various LO 
Monte Carlo models in which the hard--scattering process is described by
direct photon diagrams, namely boson--gluon fusion and QCD Compton diagrams.
The models under consideration differ in their way of describing 
the higher--order contributions to the LO process.
LEPTO~\cite{lepto} and HERWIG~\cite{herwig} use parton showers that evolve
according to the DGLAP equations.
ARIADNE~\cite{ariadne} employs the color--dipole model,
in which gluons are emitted from the color field between quark--antiquark 
pairs.
Since color dipoles radiate independently, the gluons are not ordered in 
transverse momentum, $k_T$. The linked--dipole--chain model, LDC~\cite{ldc},
implements the structure
of the CCFM equation~\cite{ccfm}, which is intended to reproduce
both DGLAP and BFKL evolution in their respective ranges of validity.
In all these models, $Q^2$ is normally used as the relevant scale.
Finally, RAPGAP~\cite{RAPGAP} 
introduces a resolved photon contribution in addition to the 
direct photon cross section 
and uses $Q^2+E^2_{T,jet}$ as the factorization scale. 
The inclusion of the resolved photon contribution
partially mimics the higher--order contributions to the direct photon
component, namely the $\log(E^2_{T,jet}/Q^2)$ terms, which are not 
included in the conventional DIS LO Monte Carlo models.
The scattering of the partons from those contained in the 
resolved photon can lead to final state
partons with a high transverse momentum in the forward direction.
Since this process was suggested to provide an explanation for the 
observed excess in forward--jet production~\cite{Jung},
the previously published forward cross 
section~\cite{ZEUS_FJets} as a function of the Bjorken scaling variable, $x$,
is compared to predictions of the RAPGAP model.
\section{Measurement}
This study is based on data taken with the ZEUS detector in 1995, 
corresponding to an integrated luminosity of 6.36~pb$^{-1}$.
As the analysis follows very closely that for the forward--jet cross 
section~\cite{ZEUS_FJets}, details about the experimental setup, event 
selection, jet finding and systematic error are not repeated here.\\

The selected DIS events were required to have a scattered electron with
a minimum energy of $E_{e'}=10$~GeV. The fractional energy transfer
by the virtual photon had to be $y>0.1$.  
The $x$~range was extended with respect to~\cite{ZEUS_FJets} from 
$4.5 \cdot 10^{-4} < x < 4.5 \cdot 10^{-2}$ to 
$2.5 \cdot 10^{-4} < x < 8.0 \cdot 10^{-2}$.
An additional cut, $Q^2>10$~GeV$^2$, was applied in order to 
be well within the DIS regime.\\

Jets were selected with a cone algorithm in the laboratory frame.
The cone radius, $R$, was chosen to be~1.0. The
transverse energy of the jets in the laboratory frame, $E_{T,jet}$,
was required to be
larger than 5~GeV and the jet pseudorapidity\footnote
{The ZEUS coordinate system is defined as right--handed 
                 with the $Z$--axis pointing in the proton beam direction, 
                 referred to as forward direction, and the
                 $X$--axis horizontal, pointing towards the center of HERA. 
                 The pseudorapidity is defined as $\eta=-\ln(\tan\frac{\theta}
                 {2})$, where the polar angle $\theta$ is taken with respect 
                 to the proton beam direction.}
range was
restricted to $\eta_{jet}<2.6$. The scaled longitudinal jet momentum 
$x_{jet}=p_{Z,jet}/p_{beam}$, where $p_{beam}=820$~GeV,
had to be larger than 0.036 to select forward jets~\cite{MUELLER91}.
Furthermore, only jets with a positive $Z$--momentum in the Breit frame
were considered, thus avoiding those jets originating from the scattered
quark at large values of $x$.
These cuts are given in 
Table~\ref{cuts}.
\vspace*{-0.2cm}
\begin{table}[htb]
\begin{center}
\begin{tabular}{c}
\hline
\hline
   $Q^2 > 10$ GeV$^{2}$         \\
   $2.5\cdot 10^{-4} < x < 8 \cdot 10^{-2}$  \\
   $y > 0.1 $                 \\
   $E_{e'} >  10$~GeV         \\
   $\eta_{jet} < 2.6$         \\
   $E_{T,jet}> 5$~GeV         \\
   $x_{jet} > 0.036$          \\
   $p_{Z,jet}(Breit) > 0$     \\
\hline
\hline
\end{tabular}
\caption{Selected kinematic region for the cross section measurement.}
\label{cuts}
\end{center}
\vspace{-0.2cm}
\end{table}
\newline
The jet cross sections presented here have been corrected 
to the hadron level for
detector acceptance and smearing
effects using the ARIADNE model, since it gave the
best description of the data~\cite{ZEUS_FJets}.  
The purity for reconstructing forward jets in the given phase space
rises from 40\% to 80\% with increasing \ETQQ , while the efficiency rises 
from 35\% to 55\%.  
For the lowest bin in \ETQQ, both purity and efficiency are around 20\%,
but here the statistical errors are large.
The factors required to  
correct the data for detector effects lie between 0.8 and 1.4
and increase as \ETQQ\ increases.
\section{Results}
\begin{figure}[htb]
\begin{center}
\psfig{figure=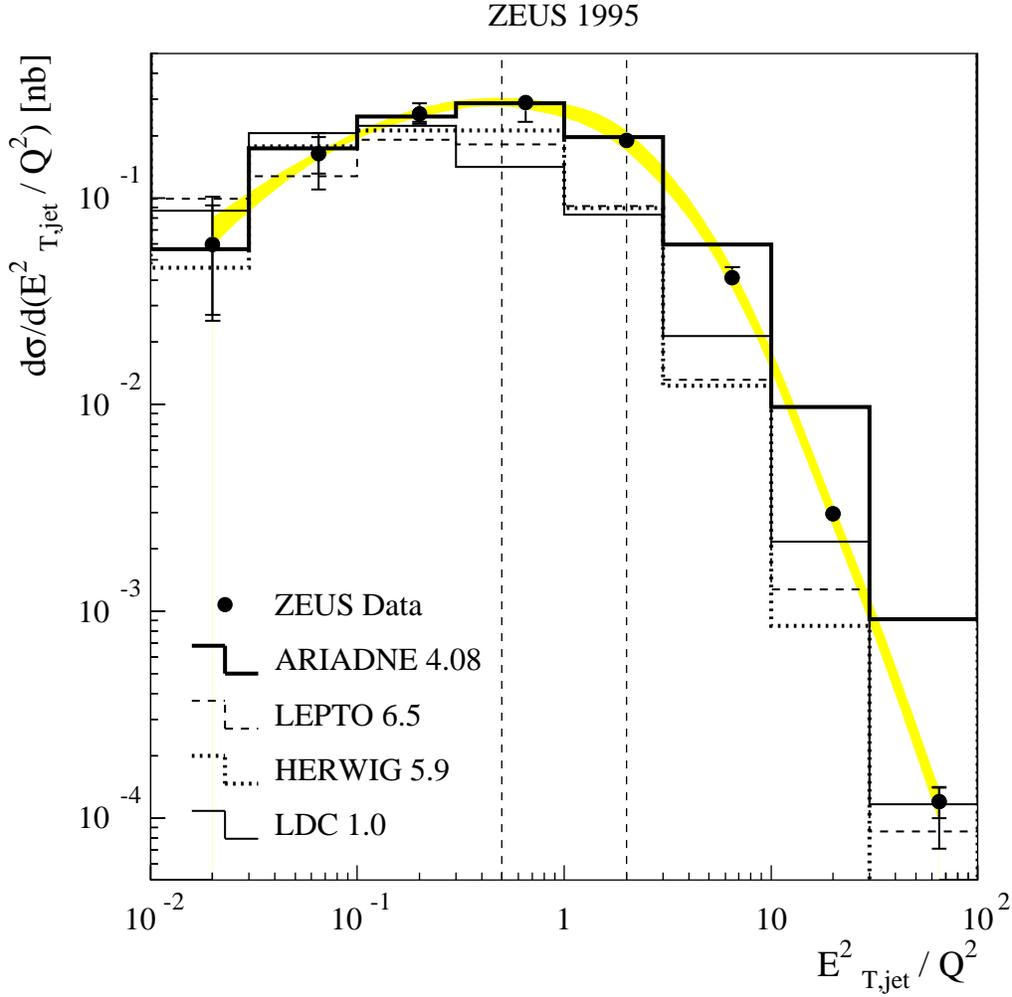,width=15.cm} 
\end{center}
\vspace*{-1cm}
\caption{Forward--jet cross sections as a function of \ETQQ. 
         ZEUS data are shown as points with 
         the inner error bars indicating the statistical errors. The outer 
         error bars give the statistical and systematic errors added in 
         quadrature.
         The shaded band corresponds to the uncertainty from the
         energy scale of the calorimeter.
         The Monte Carlo predictions from ARIADNE (thick, full line),
         LEPTO (dashed line), HERWIG (dotted line) and LDC (thin, full line)
         are shown for comparison.
         The vertical dashed lines indicate the region used for the
         previous forward cross section measurement~\cite{ZEUS_FJets}.}
\label{fig:ETQQ}
\end{figure}

The forward--jet cross section is presented in Fig.~\ref{fig:ETQQ}
as a function of \ETQQ.
The numerical values are given in Table~\ref{values}. The treatment of the
systematic errors closely follows the published results~\cite{ZEUS_FJets}
and leads to errors of similar size. The shaded band corresponds to
the uncertainty coming from the energy scale of the calorimeter.\\

\begin{figure}[htb]
\begin{center}
\epsfig{figure=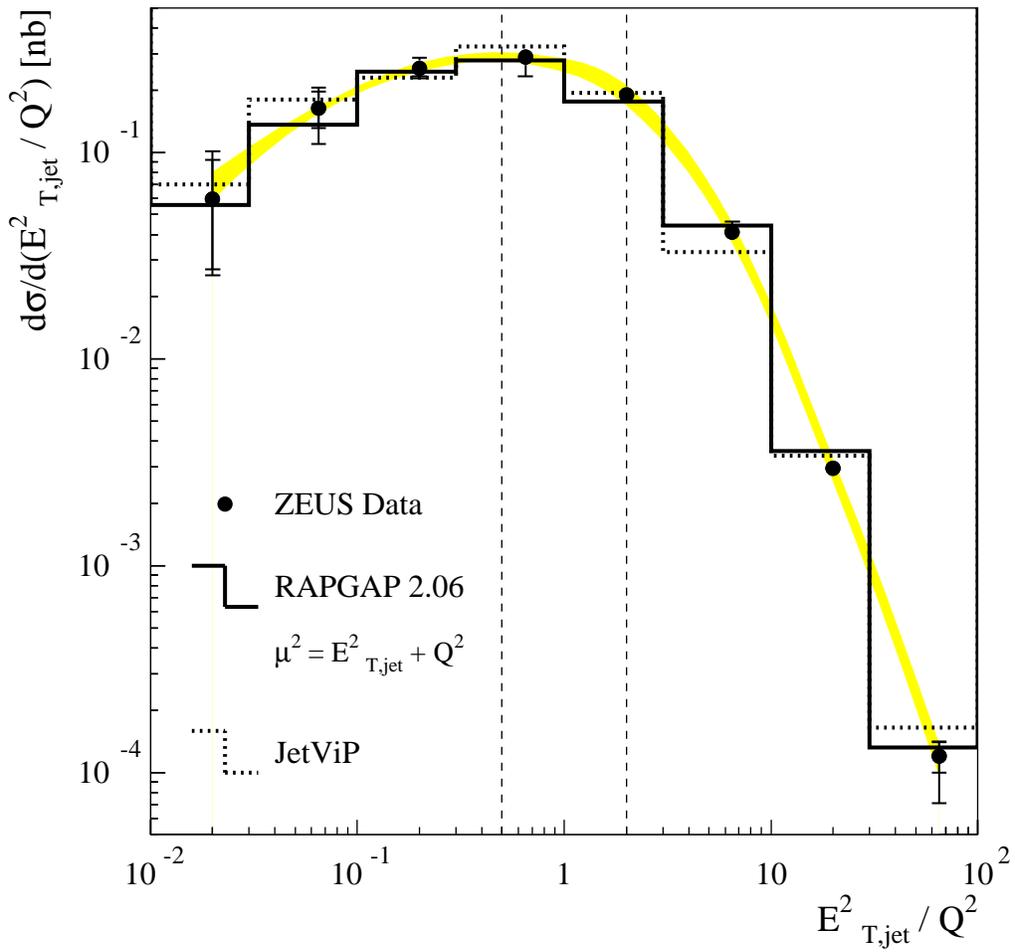,width=15.cm} 
\end{center}
\vspace*{-1cm}
\caption{Forward--jet cross section as a function of \ETQQ.
         Data points, error bars and the error band are the same as in 
         Fig.~\ref{fig:ETQQ}.
         The data are compared to the RAPGAP Monte Carlo model with
         direct and resolved contributions (full histogram).
         The results of the NLO calculation JetViP are
         shown as the dotted histogram~\cite{poetter}.}
\label{fig:rap_noptq}
\end{figure}

Predictions from different LO Monte Carlo models are shown
in Fig.~\ref{fig:ETQQ} and Fig.~\ref{fig:rap_noptq}.
Three regions are distinguished, 
separated by the dashed vertical lines.
In the region where $Q^2 \gg E^2_{T,jet}$, all the models describe the
data reasonably well.\\

In the regime
$Q^2 \approx E^2_{T,jet}$, only ARIADNE~4.08 and RAPGAP~2.06 reproduce 
the measured distributions.
In RAPGAP the resolved component of the virtual photon is modeled using
the SaS-2D parametrization for the parton distribution 
function (pdf) of the photon~\cite{Schuler}, which in this $Q^2$ range
evolves as $\sim \log(E_T^2/Q^2)$. The factorization scale
has been set to $\mu^{2}=E^{2}_{T,jet}+Q^2$.\\

\begin{figure}[htb]
\begin{center}
\epsfig{figure=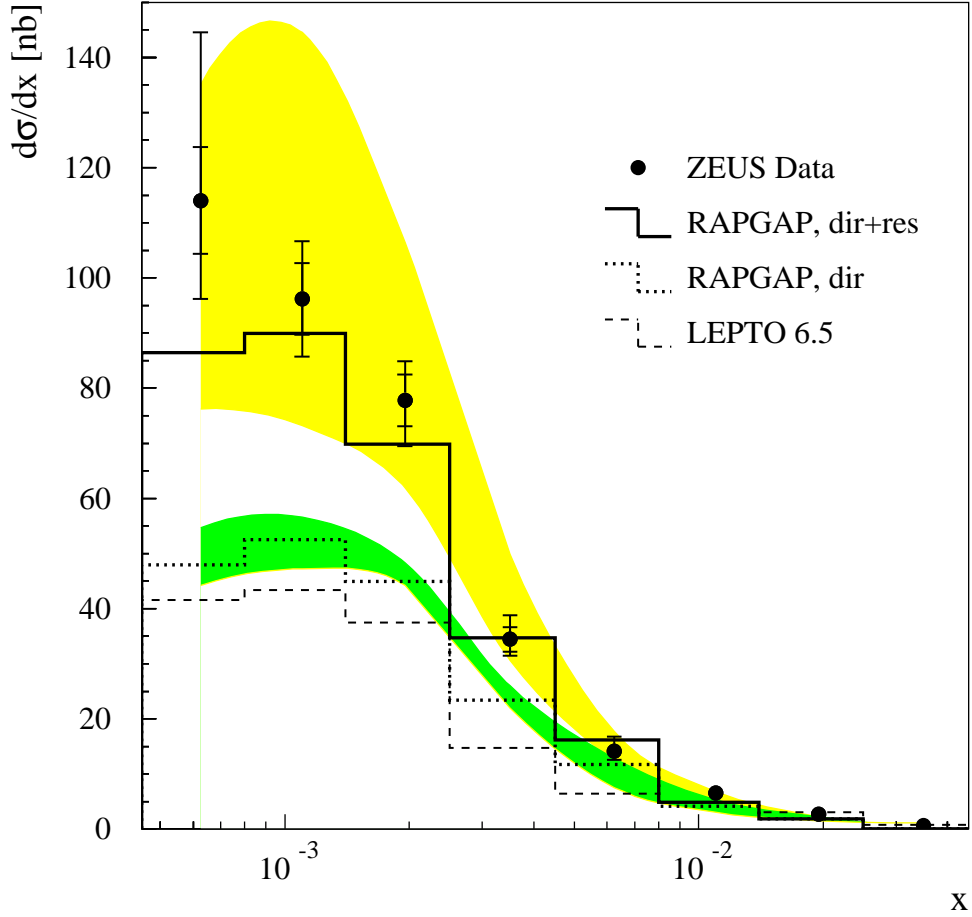,width=15.cm} 
\end{center}
\vspace*{-1.cm}
\caption{Forward--jet cross section as a function of $x$, in the region
         $ 0.5 < \ETQQ < 2.0$.
         The ZEUS data, shown as points, are compared to the RAPGAP Monte 
         Carlo model with direct and resolved photon contributions (full line) 
         and to LEPTO (dashed line). The ARIADNE prediction (not shown) 
         is very
         similar to RAPGAP. The direct contribution of RAPGAP
         is indicated by the dotted line.
         The factorization scale in RAPGAP has been varied from 
         $\mu^{2}=E^{2}_{T,jet}/2+Q^2$ to $\mu^{2}=4 \cdot E^{2}_{T,jet}+ Q^2$.
         This corresponds to the light shaded band which refers to
         the full 
         RAPGAP histogram (sum of the resolved and direct components) and 
         to the dark shaded band which refers to the direct RAPGAP histogram.
         The histograms correspond to the nominal scale 
         $\mu^{2}=E^{2}_{T,jet}+Q^{2}$.}
\label{fig:hadron_rap}
\end{figure}

In Fig.~\ref{fig:hadron_rap} the $x$--dependence in this regime is compared
with RAPGAP, using the cuts
$0.5<\ETQQ<2.0$ and $4.5\cdot 10^{-4}<x<4.5\cdot 10^{-2}$~\cite{ZEUS_FJets}.
RAPGAP gives a good description of the cross section. The contribution
of the direct photon component is indicated separately. As expected, 
it matches the LEPTO prediction.\\

For $Q^2 \ll E^2_{T,jet}$, none of these models, except RAPGAP, 
reproduces the data. In particular ARIADNE overshoots the data by up to an 
order of magnitude at the upper limit of the displayed range. 
The other models, 
LEPTO~6.5, HERWIG~5.9 and LDC~1.0, lie far below the data.
These comparisons using corrected cross sections are similar to
those made previously~\cite{ZEUS_FJets}, using the uncorrected
distributions.
The same data are shown in Fig.~\ref{fig:rap_noptq} 
together with the prediction of the RAPGAP Monte Carlo model,
which describes the data well over the full range of 
\ETQQ. \\

Recently, the parton level NLO calculation JetViP~\cite{jetvip} has become 
available, to which our data can also be compared, with the proviso that
the hadronization corrections are model--dependent and are of the order 
of up to 20\%.
JetViP sums contributions from the direct and 
resolved virtual photon and uses the SaS-1D photon pdf~\cite{Schuler}.
For the first three bins in Fig.~\ref{fig:rap_noptq}
only the direct contribution has been taken into account, since $Q^2$ is
large enough ($Q^2>83$~GeV$^2$) that the resolved component can be
neglected.
The renormalization and factorization scales 
have been set to $E_{T,jet}^2+Q^2$~\cite{poetter}. 
The agreement over the full range of 
\ETQQ\ is good. The $x$ dependence of the cross section in the
range $0.5 < \ETQQ\ < 2.0$ has also been calculated with 
JetViP~\cite{kramerpoetter}
and good agreement was found. The fact that only RAPGAP and JetViP
describe the data implies that a resolved photon component is
necessary for $\ETQQ > 1$.\\

The necessity of a resolved photon component in a DIS process 
has also been discussed by the H1 collaboration 
in the context of dijet production in a
$Q^2$ range of 5 to 100~GeV$^2$~\cite{h1_dijets},
where the measured dijet cross section could only be described with the
inclusion of the resolved component.\\

In comparing the performance of RAPGAP and JetViP it should be noted that
while they both agree with the data, their predictive power
is limited. 
On the one hand both RAPGAP and JetViP use the SaS photon pdf, which for
$Q^2>0$ is not very well constrained by experimental data.
On the other hand there is a large 
variation of the results when the factorization scale is varied, as shown by
the light shaded band in Fig.~\ref{fig:hadron_rap} for RAPGAP.
A similar effect is seen for JetViP~\cite{kramerpoetter}.
\section{Summary}
The cross sections for 
forward--jet production over a wide range of \ETQQ\ have been
compared to different Monte Carlo models. 
All leading--order Monte Carlo models tested here 
give a good description of the region in which $E^2_{T,jet} \ll Q^2$.
However, only those models which include non--$k_T$--ordered gluon emissions, 
or contributions from a resolved photon, reproduce the 
$E^2_{T,jet} \approx Q^2$ region.
The full range of \ETQQ\ can be described only
by the RAPGAP model
and the JetViP NLO QCD calculation, both of which include
a resolved photon contribution.
The forward--jet differential 
cross section, as a function of~$x$~\cite{ZEUS_FJets}, 
is also well reproduced by RAPGAP and JetViP. However, the 
large dependence of its predictions on the factorization scale 
diminishes the significance of this agreement.
\section*{Acknowledgements}
We thank the DESY directorate for their strong support and encouragement.
The remarkable achievements of the HERA machine group were essential for
the successful completion of this work and are gratefully acknowledged.
We also thank G.~Kramer for useful discussions and B.~P\"otter for 
providing the JetViP calculation.

\newpage

\begin{table}[htb]
\renewcommand{\arraystretch}{1.9}
\begin{center}
\begin{tabular}{|r@{ -- }l||r@{$\,\pm\,$}l@{$\,^{+}_{-}\,$}l|r@{, +}l|}
\hline
\multicolumn{2}{|c||}{$E_{T,jet}^{2}/Q^{2}$} & 
\multicolumn{3}{c|}
{$\frac{d\sigma}{d(E_{T}^{2}/Q^{2})} 
\pm{\rm stat.}\pm{\rm syst.}~[{\rm pb}]$} & 
\multicolumn{2}{c|}{syst.\ $E_{CAL}$--scale [pb]}\\
\hline \hline
0.01 & 0.03 &\hspace*{0.8cm} 59.5  & 32.3  & $^{26.5}_{10.7}$   &\hspace{0.6cm} ($-$0.1   & 19.5) \\
0.03 & 0.1  & 164   & 33    & $^{26}_{43}$       & ($-$9     & 2) \\
0.1  & 0.3  & 255   & 22    & $^{24}_{17}$       & ($-$6     & 13) \\
0.3  & 1.0  & 288   & 12    & $^{12}_{53}$       & ($-$16    & 9) \\
1.0  & 3.0  & 190   & 6     & $^{2}_{7}$         & ($-$19    & 18) \\
3.0  & 10.  & 41.2  & 1.4   & $^{4.8}_{0.9}$     & ($-$4.0   & 3.5) \\
10.  & 30.  & 2.95  & 0.19  & $^{0.13}_{0.13}$   & ($-$0.33  & 0.27) \\
30.  & 100. & 0.120 & 0.020 & $^{0.007}_{0.045}$ & ($-$0.021 & 0.014) \\
\hline
\end{tabular}
\caption{Cross section values and errors for the corrected data
in bins of \ETQQ.
The last column shows the systematic errors due to the 
energy scale uncertainty of the calorimeter, which is not included in the 
central column. The table refers to the points shown in Fig.~\ref{fig:ETQQ}. 
The phase space under investigation is defined by the cuts: \etajet$<2.6$, 
\xjet$>0.036$, $E_{T,jet}>5$~GeV, $E_{e'} > 10$~GeV, $y>0.1$, 
\qq$>10$~GeV$^2$, $p_{Z,jet}(Breit)>0$ 
and~$2.5\cdot10^{-4}<x<8.0\cdot10^{-2}$.}
\label{values}
\end{center}
\end{table}
\end{document}